\documentclass [11pt]{article}
\usepackage {graphicx}
\usepackage{amsfonts}

\def\Z{{\mathbb Z}}
\def\R{{\mathbb R}}
\def\Ker{\mathrm{Ker}\,}
\def\Im{\mathrm{Im}\,}

\def\rank{\mathrm{rank}\,}
\def\const{\mathrm{const}\,}
\def\GL{\mathrm{GL}\,}

\begin{document}

\title{Numerical analysis of topological characteristics of three-dimensional
geological models of oil and gas fields
\footnote{The work is supported by RFBR (project 11-01-12106-ofi-m-2011) and the grants of President of Russian Federation (projects MD-249.2011.1 and NSh-544.2012.1).}}
\author{Ya.V. Bazaikin
\footnote{Sobolev Institute of Mathematics, 630090 Novosibirsk, Russia; e-mail: bazaikin@math.nsc.ru}
\and
V.A. Baikov
\footnote{Ufa State Aviation Technical University, 450025 Ufa , Russia; e-mail:baikov@ufanipi.ru}
\and
I.A. Taimanov
\footnote{Sobolev Institute of Mathematics, 630090 Novosibirsk, Russia; e-mail: taimanov@math.nsc.ru}
\and
A.A. Yakovlev
\footnote{Institute of Mathematics, 450008 Ufa, Russia; e-mail: yakovlevandrey@yandex.ru}}
\date{}

\maketitle

\begin{abstract}
We discuss the study of topological characteristics of random fields that are used for
numerical simulation of oil and gas reservoirs and numerical algorithms, for computing such
characteristics, for which we demonstrate results of their applications.

\end{abstract}

\section{Introduction}

For the efficient extraction of oil (or gas) from oil and gas reservoirs
modern technology is needed to monitor the development of
fields and, in particular, the methods of geological and hydrodynamic
modeling and geosteering. Now for re\-pro\-duction of the real structure formation
there are used probabilistic methods of
digital geological modeling \cite{1}, which are as follows:

An oil (gas)-bearing bed is discretized, i.e. is represented by a grid, i.e. a cover by
disjoint cells which in practice often consists of
parallelepipeds of $50$ m horizontal breadth and $0.4$ m vertical depth.

Further, to each cell, through stochastic modeling, based on
conceptual structure formation and statistical evaluation of the input
geological infor\-ma\-tion, there are attributed its capacitive--filtration properties
such as porosity, permeability, compressibility, etc.

In the development of a reservoir to control the flow of a fluid in it
there are used hydrodynamic simulations which are software
products for numerical solving multiphase filtration equations for which there  no
analytical solutions are available. Numerical calculation is
extremely resource consumptive and
this factor is decisive in modeling with more
cells, and, in particular, determines the size of the integrated (in the
``upscaling'') computational cells, as well as the need for further
zeroing some of them. Moreover, for adequate reproduction of the flow
it is obviously necessary to use averaging filter equations and determination
of effective capacitive--filtration characteristics of the design grid cell.

Hence, the main purpose of a digital geological modeling in the oil
industry is the creation of a geological reservoir model by
regard to its geometric characteristics and by determining reservoir
properties. Therewith it must be assumed that the accuracy of the resulting model
is determined by the hydro\-dy\-na\-mic model of the flow of a fluid in the reservoir (we want to emphasize
uselessness of excessive detail.) In practice the process of creation of a geological and hydrodynamic
formation model is consecutive and often independent, i.e., first by
geologists and then by developers. At one stage some design principles
are used and on the other - the other, and they often contradict each other.
Therefore, it is necessary at an early (conceptual) stage of a
geological modeling to take into account the development data, and of course,
they are unavoidable in the construction of a digital geological model. The
natural question arises: how to link geology and field development. This is a
central issue that motivates our research. Unfortunately there is no explicit answer
to this question But it is clear that we cannot do without a list of characteristics
of random fields (digital geological models).

We recall some geometrical characteristics of three-dimensional digital
geological and reservoir simulation models: the distribution of the lengths of streamlines \cite {2},
the decline rate of a well discharge at a constant depression \cite {3}, and the fractal
dimension of the model and its percolation properties \cite {4}. The present work is
devoted to the study of topological characteristics of random fields
(geological and hydrodynamic models) and their relation to the solution of filtration equations
(the data of field development).

The problem of efficient computation of numerical characteristics describing
the topological structure of complex geometrical objects is a subject of
computational topology that is being actively
developing in recent years \cite {5,6,7}. In this case,
often the objects under study are given as a series of nested into
topological spaces (i.e., by filtration), and
we have to trace how the topological structure of
subspaces changes in the process of getting the original space. In topology, and
more precisely in Morse theory, we have to deal with the filtration of topological
spaces, which arises when considering the excursion sets of a
function $f$; i.e., the sets of the form $ \{f \ge c_0 = \const \} $. In this situation,
we have to follow the change of the topological structure of the excursion set
when $c_0$ is changed.

In applications $f$ by itself can have a probabilistic nature, and,
so, there appears a problem of statistical evaluation of the impact of the excursion level
on the topological characteristics of excursions;

{\sl there is a problem of distinguishing topological invariants that
are persistent under small perturbations of the excursion level.}

A useful tool for the study of these and other related problems are
homology and Betti numbers, and for investigation of  the dependence of topological
properties of the excursion set on its level one can use persistent
homology and persistent Betti numbers. Roughly speaking, the persistent homology
estimate the portion of homology that ``survives'' for a given change
of the level of the function. A detailed account of persistent homology and applications
can be found in \cite{8,9,10,12,16,11,14,15,17}.

When modeling the formation there naturally arises the permeability function
defined by its values in each grid cell. The excursion set
$\{f \ge c_0 \} $ of this function is a three-dimensional body
modeling a reservoir for a given threshold of permeability. We note that this
definition of a reservoir as the excursion set of the permeability function
carries certain dangers in simulation, since it is difficult to clearly
specify the excursion level in which we distinguish permeable and
impervious areas, especially when we consider the probabilistic nature of this
function. Thus, the method of comparison of implementations must be stable
under fluctuations of excursion levels that lead us to use
persistent topological characteristics. We note that usually various
applications of persistent homology are due to resistance to noise at
changing objects.

In conclusion, we note that we study topological characteristics of
realizations of a random field: calculation is made after the choice of an implementation
and the excursion level. It is a reasonable problem  of computing
characteristics with taking into account the probabilistic nature of the object, i.e.
estimation of the topological characteristics of excursions sets of the random field that models
the structure formation. Problems of the kind were studied in \cite{18}.
An important problem that arises is the formulation of filtration equations for a random field
and the relation of solutions to these equations to the characteristics of the random field  and that
is a subject for further research.

\section{Computation of Betti numbers}

\begin{figure}[htbp]
\centerline{\includegraphics[width=5.00in,height=2.82in]{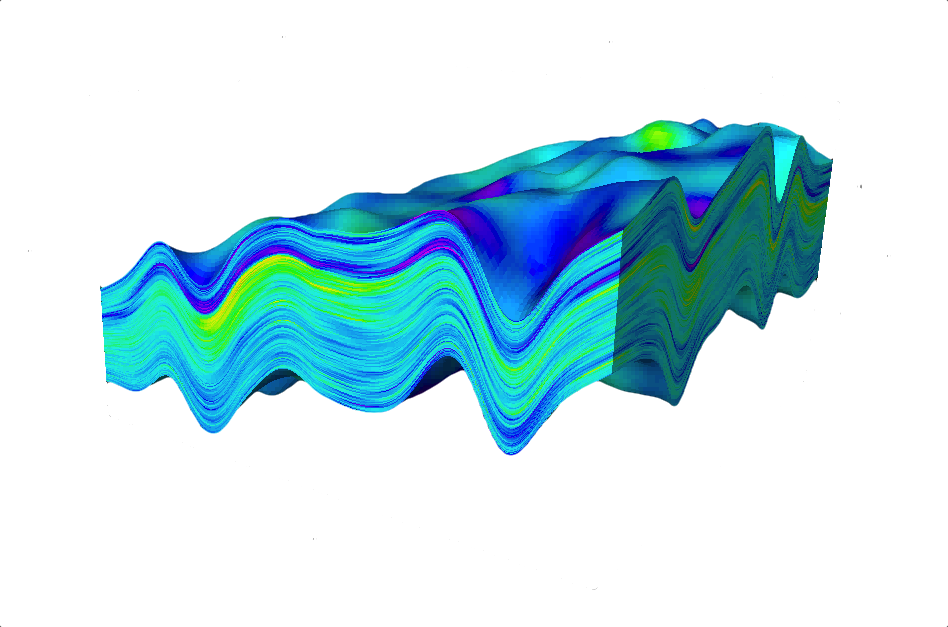}}
\centerline{\includegraphics[width=2.50in,height=0.3in]{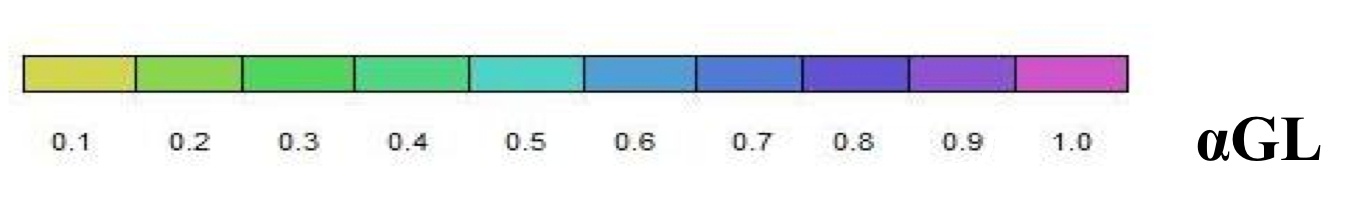}}
\caption{Realization of an oil reservoir by SPECTRAL.}
\label{fig1}
\end{figure}

\begin{figure}[htbp]
\centerline{\includegraphics[width=5in,height=2.82in]{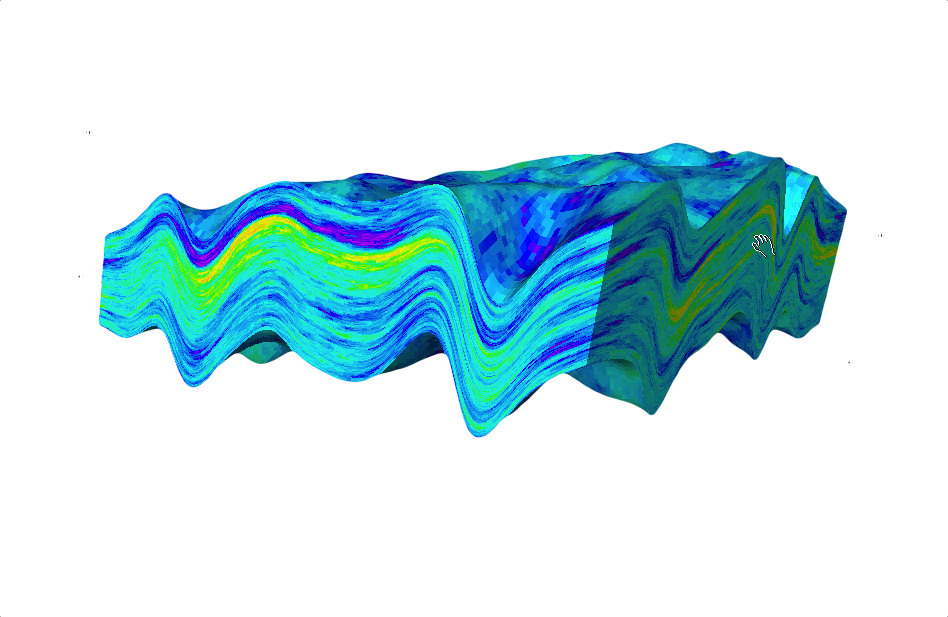}}
\centerline{\includegraphics[width=2.50in,height=0.3in]{Scale-eng.jpg}}
\caption{Realization of an oil reservoir by SGS.}
\label{fig2}
\end{figure}

In \cite{19} there is presented a numerical algorithm for computing topological inva\-ri\-ants of three-dimensional bodies
by using a discrete version of Morse theory.
These invariants are the Betti numbers  $b_0 ,b_1$, and $b_2$, i.e. the
numbers of connected components, of independent one-dimensional cycles and of ``voids''
in the body. These characteristics  have clear interpretations in terms of the perme\-abi\-lity of a reser\-voir:
the connectedness and compartmentalization of a reservoir play primary roles in problems of its development.
In Fig. \ref{fig1} and Fig. \ref{fig2} we present different realization of the
same reservoir that are obtained by different methods called SGS and SPECTRAL.
The SGS method (a succes\-si\-ve Gauss simulation \cite{20}) is widely
accepted and is based on the assumption that geophysical fields are stationary both laterally and vertically.
The method SPECTRAL was presented in
\cite{21,22}, and its main difference consists in the following representation of a
geophysical field:
$$
\xi (x,y,h) = \sum\limits_k {a_k } (x,y)L_k (h),
$$
where $x$ and $y$ are the lateral variables, $h$ is the vertical variable, $L_k (h)$ are the Legendre polynomials,
and the random processes $a_k (x,y)$ are assumed stationary.


\begin{figure}[htbp]
\centerline{\includegraphics[width=5.00in,height=2.82in]{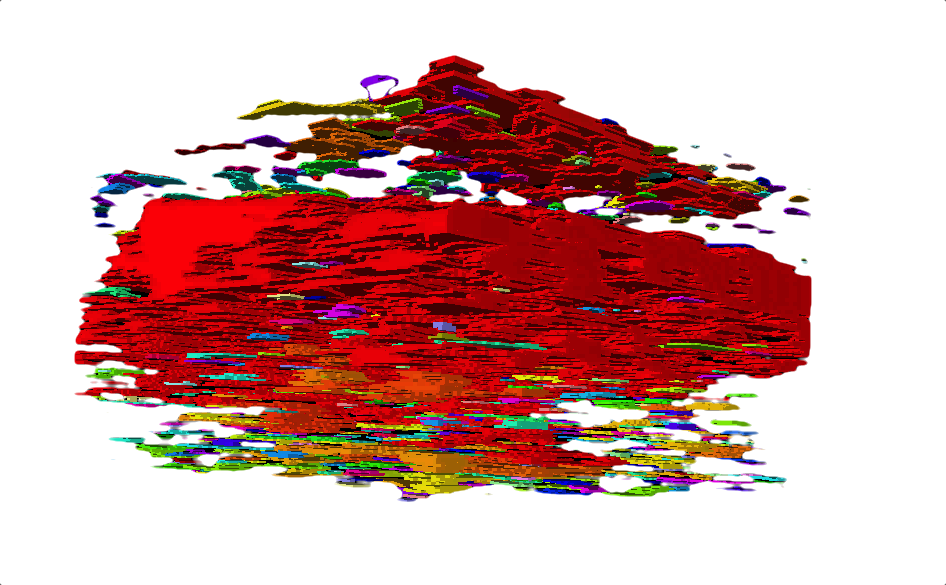}}
\caption{Excursion of a realization of oil reservoir obtained by SPECTRAL.}
\label{fig3}
\end{figure}

\begin{figure}[htbp]
\centerline{\includegraphics[width=5.00in,height=2.82in]{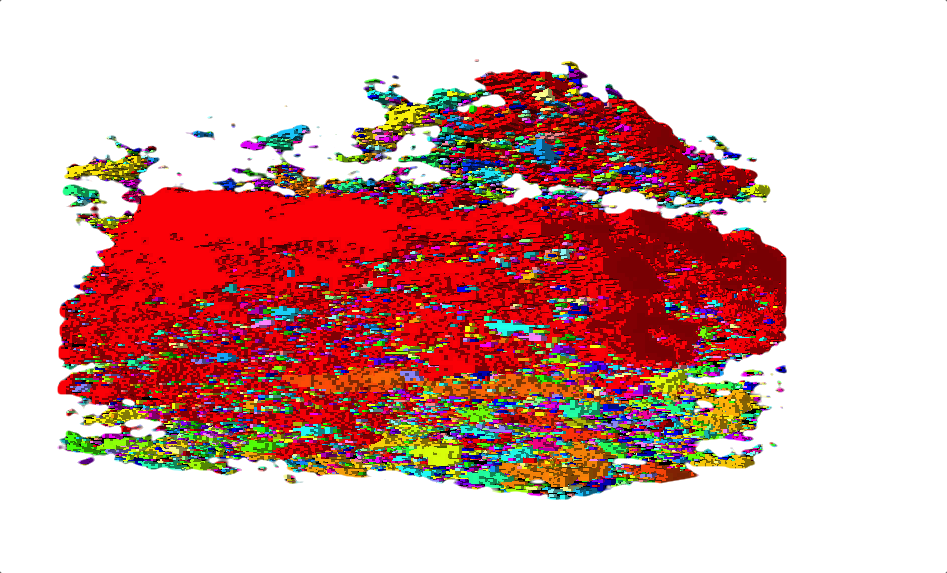}}
\caption{Excursion of a realization of oil reservoir obtained by SGS.}
\label{fig4}
\end{figure}

Here $\GL$ (the gamma logging) is the natural radioactivity of formation, and a reservoir is modeled
as the excursion set $\{\alpha \GL \le \const\}$ of the function
$$
\alpha \GL = \frac{\GL - \GL_{\min} }{\GL_{\max} - \GL_{\min} },
$$
defined on the cube of size $120 \times 120 \times 490$. We note that we have the reverse
inequality in the definition of an excursion because the permeability of a formation is inverse to its
radioactivity. In Fig. \ref{fig3} and Fig. \ref{fig4} there are displayed the excursion sets
$\{\alpha \Gamma K \le 0.6\}$ for the two realizations given in Fig. \ref{fig1} and Fig. \ref{fig2}.
Different colors correspond to different connected components.


\begin{figure}[htbp]
\centerline{\includegraphics[width=6.12in,height=2.59in]{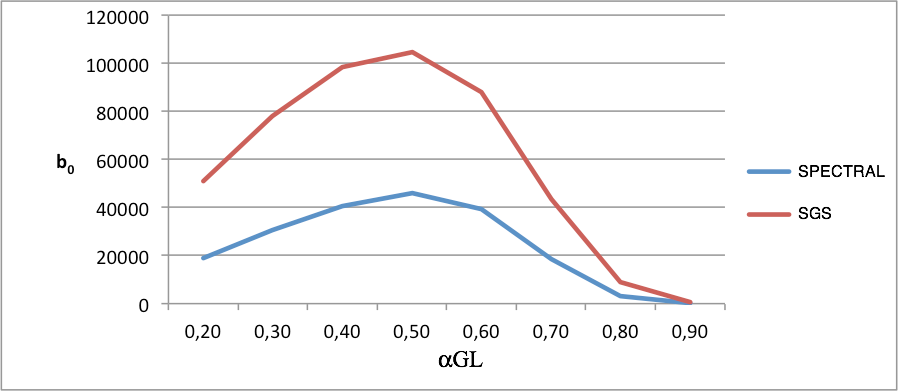}}
\caption{The $0$-th Betti number.}
\label{fig5}
\end{figure}

\begin{figure}[htbp]
\centerline{\includegraphics[width=5.97in,height=2.30in]{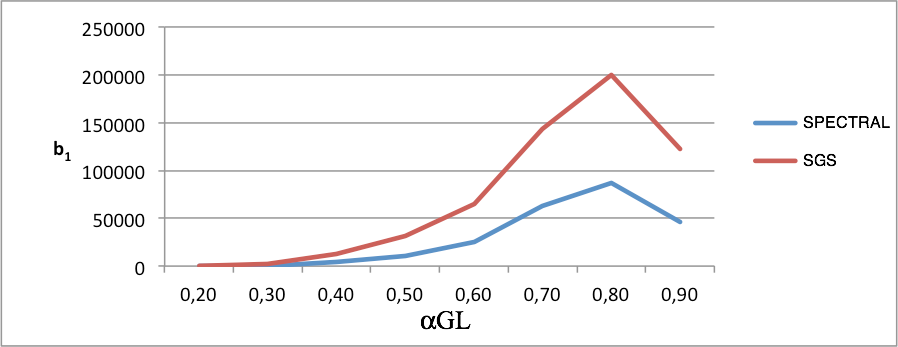}}
\caption{The $1$-st Betti number.}
\label{fig6}
\end{figure}

\begin{figure}[htbp]
\centerline{\includegraphics[width=5.97in,height=2.40in]{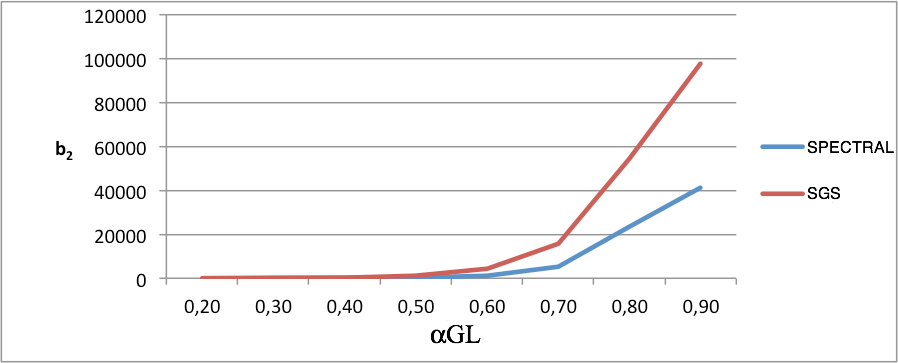}}
\caption{The $2$-nd Betti number.}
\label{fig7}
\end{figure}

\begin{figure}[htbp]
\centerline{\includegraphics[width=5.97in,height=2.38in]{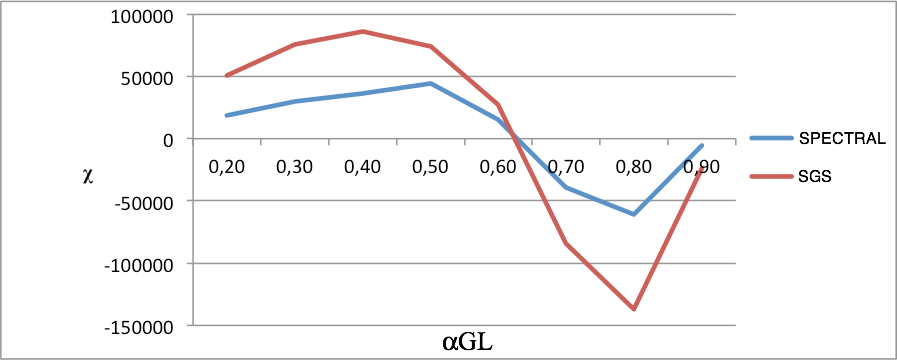}}
\caption{The Euler characteristic.}
\label{fig8}
\end{figure}

For every realization the Betti numbers and the Euler characteristic
$\chi = b_0 - b_1 + b_2$ are computed. Table \ref{tab1} demonstrates the dependence of
the Betti numbers and the Euler characteristic on the excursion level.
For every excursion level we give results of computing the topological characteristics of
two different models of the same reservoir that are obtained by
the method SPECTRAL (the upper line) and by the SGS method (the lower line).
The last column contains the duration of computations on the processor
Intel\textregistered Core\texttrademark i7 3.33GHz. In Fig. \ref{fig5}--\ref{fig8}
one may find graphs of different characteristics of excursion for both methods.
We note that Betti numbers may distinguish the models of reservoirs obtained by the
different methods
of geo\-sto\-chas\-tic modeling from the same geophysical data.

\begin{table}[htbp]
\begin{tabular}
{|p{30pt}|p{40pt}|p{40pt}|p{40pt}|p{40pt}|p{60pt}|}
\hline
$\alpha \mbox{Г}\mbox{К}$&
b$_{0}$ &
b$_{1}$ &
b$_{2}$ &
$\chi $&
Time (hr:min:sec)  \\
\hline
\raisebox{-1.50ex}[0cm][0cm]{0.2 }&
19085 &
72 &
0 &
19013 &
00:00:07  \\
\cline{2-6}
 &
50874 &
252 &
3 &
50625 &
00:00:12  \\
\hline
\raisebox{-1.50ex}[0cm][0cm]{0.3 }&
30647 &
567 &
3 &
30083 &
00:00:24  \\
\cline{2-6}
 &
78291 &
2634 &
29 &
75686 &
00:00:41  \\
\hline
\raisebox{-1.50ex}[0cm][0cm]{0.4 }&
40420 &
3977 &
34 &
36446 &
00:00:52  \\
\cline{2-6}
 &
98672 &
13162 &
298 &
85808 &
00:03:31  \\
\hline
\raisebox{-1.50ex}[0cm][0cm]{0.5 }&
46029 &
10934 &
196 &
44291 &
00:02:34  \\
\cline{2-6}
 &
104647 &
31758 &
1287 &
74176 &
00:13:58  \\
\hline
\raisebox{-1.50ex}[0cm][0cm]{0.6 }&
39377 &
24800 &
1167 &
15744 &
00:08:15  \\
\cline{2-6}
 &
88255 &
65012 &
4471 &
27714 &
00:37:23  \\
\hline
\raisebox{-1.50ex}[0cm][0cm]{0.7 }&
18563 &
62533 &
5136 &
-38834 &
00:33:20  \\
\cline{2-6}
 &
43630 &
143720 &
15785 &
-84305 &
01:41:54  \\
\hline
\raisebox{-1.50ex}[0cm][0cm]{0.8 }&
3106 &
87319 &
23308 &
-60905 &
00:29:57  \\
\cline{2-6}
 &
8854 &
200174 &
54334 &
-136986 &
01:37:07  \\
\hline
\raisebox{-1.50ex}[0cm][0cm]{0.9 }&
174 &
46653 &
41312 &
-5167 &
00:06:28  \\
\cline{2-6}
 &
577 &
122147 &
97657 &
-23913 &
00:20:43  \\
\hline
\raisebox{-1.50ex}[0cm][0cm]{1.0 }&
4 &
15318 &
31022 &
15708 &
00:01:47  \\
\cline{2-6}
 &
26 &
38288 &
76722 &
38460 &
00:01:47  \\
\hline
\end{tabular}
\caption{The Betti numbers and the Euler characteristic.}
\label{tab1}
\end{table}

{\sc Remark.} A computation of topological characteristics demonstrates the difference between
the methods of geostochastical modeling, i.e. between SGS and SPECTRAL: the Betti numbers
for different models of the same reservoir may vary upto $2-6$ times.

\section{Persistent homology}

Rigorous exposition is given, for instance, in \cite{23,24} of cell complexes and the basic ideas and
constructions of Morse theory that we use in the sequel.

For computing the topological characteristics of a space it is
convenient to represent the space as a union of elementary ``bricks,'' i.e. cells, which
are ``correctly'' glued to each other. The resulted space is called
a {\it cell complex}. By a $0$-dimensional cell one means a point,
and a union of finitely many $0$-dimensional cells forms the $0$-th skeleton
$X^0$ of a cell complex $X$. Let us consider a family of $1$-dimensional cells, i.e. intervals
glued to $X$ so that the ends of the intervals are identified with certain
$0$-cells. The resulted space would be the $1$-dimensional skeleton $X^1$.
We construct the complex $X$ by successively gluing $i$-dimensional discs to
$(i-1)$-dimensional sceleta $X^{i-1}$.

For our purposes it is enough to use cubic complexes, i.e. such cell complexes that all
$i$-cells are $i$-dimensional cubes are glued to $X^{i-1}$ as follows: every boundary face of a $i$-cell is
an $(i-1)$-dimensional cube which is identified with some $(i-1)$ cube from $X^{i-1}$.

Let us consider the filtration of a cell complex $X$ by cell subcomplexes:
$$
\emptyset = X_0 \subset X_1 \subset \ldots \subset X_n = X.
$$

We consider homology with coefficients
in the residue group $\Z_2$.
The filtration defines the chain of homomorphisms of the homology groups
$H_q^p = H_q (X_p )$:
$$
0 = H_q^0 \to H_q^1 \to \ldots \to H_q^n \to H_q^{n + 1} = 0
$$
for every $q \ge 0$. The compositions of successive homomorphisms from the chains give rise
to the homomorphisms
$$
f_q^{i,j}: H_q^i \to H_q^j.
$$

By definition, {\it the persistent homology groups of dimension $q$}
are the groups
$$
H_q^{i,j} = \Im f_q^{i,j}  \ \ \ \mbox{for $0 \le i \le j \le n + 1$.}
$$
Respectively by $q$-th persistent Betti numbers we mean the ranks of
the persistent homology groups: $b_q^{i,j} = \rank H_q^{i,j}$. In particular,
$H_q^{i,i} = H_q^i $.

Let us fix $q$ and choose a basis
$\{e_1^i ,e_2^i,\ldots ,e_{m_i }^i \}$ for $H_q^i $ such that for every $1 \le k \le m_i
$, $f_q^{i,i + 1} (e_k^i ) \in \{0,e_1^{i + 1} ,\ldots ,e_{m_{i + 1} }^{i +
1} \}$ for every $1 \le k \le m_i$ and $f_q^{i,i + 1} (e_k^i ) = f_q^{i,i + 1} (e_{k^\prime }^i ), k
\ne k^\prime $т if and only if $f_q^{i,i + 1} (e_k^i ) = 0$.
Hence $H_q^{i,i + 1}$ consists of such elements $e_k^i $ that do not vanish, i.e. survive.
Respectively the persistent homology group $H_q^{i,j} $ consists of elements $e_k^i \in
H_q^i $ that survive up to $H_q^j $.

There is a useful graphical representation for persistent homology that is called a {\it barcode}
\cite{9,10,11}. Namely, given the dimension $q$, let us consider a basic element
$e_k^i$ that is not an image of any element from $H^{i-1}_q$. Then here exists a minimal value $j \ge i$
such that $f_q^{i,j} (e_k^i ) = 0$. Then we correspond to $e_k^i$ the interval $(i,j)$.
A disjoint union of all such intervals is usually portrayed on the two-plane by intervals parallel to the
$Ox$ axis and forms the $q$-barcode. It gives a visual representation for changing of topology of
$X_i $ with increasing of $i$.

\section{Computation of homology}

In this section we demonstrate the main ideas of the numerical algorithm for computing
the Betti numbers of three-dimensional bodies which is presented in \cite{19}
by using an example of computing the persistent $0$- and $2$-homology
and present some results of computations.

Let us consider some cubic domain, in the Euclidean space,
$$
K = [0,N]\times [0,N]\times [0,N] \subset \R^3,
$$
with some natural $N$. By an elementary interval $I \subset \R$
we mean a set of the form
$$
I = [l,l + 1],
$$
where $l$ is some natural number. Analogously we define natural square
$$
Q = I_1 \times I_2 \subset \R^2,
$$
and elementary cube
$$
C = I_1 \times I_2 \times I_3 \subset \R^3,
$$
where $I_k, k=1,2,3$ are elementary intervals. Hence the domain $K$ consists of
elementary cubes.

Let $M_1,\dots,M_n$ be $3$-dimensional bodies formed by elementary cubes and lying inside $K$.
We assume that every $M_i$ is the excursion set $\{f \geq c_i\}$ for some continuous function
$f$ defined on elementary cubes from $K$ and $c_i > c_j$ for  $i < j$.
Hence we have the filtration
$$
M_1 \subset M_2 \subset \dots \subset M_n.
$$
Variation of an excursion level from $c_1$ to $c_n$ results in
variation of the topology of the excursion sets and that may be described in terms of the persistent homology
$H_\ast ^i = H_\ast (M_i )$.

By applying, if need be, the preprocessing of $M$ \cite{19}, we
assume that two elementary cubes from $M$ may not touch each other only at a vertex or along an edge.
In applications that means that oil may pass from one cell to another only through
a common $2$-dimensional face and there is no oil passing through common
vertices and edges.

\begin{figure}[htbp]
\centerline{\includegraphics[width=2.00in,height=2.00in]{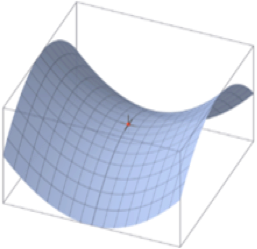}
\includegraphics[width=2.00in,height=2.00in]{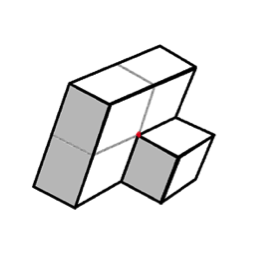}}
\caption{The saddle and its discrete analog.}
\label{fig9}
\end{figure}

\begin{figure}[htbp]
\centerline{\includegraphics[width=2.00in,height=2.00in]{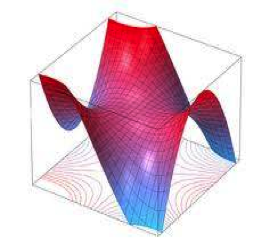}
\includegraphics[width=2.00in,height=2.00in]{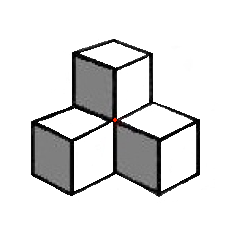}}
\caption{The ``monkey saddle'' and its discrete analog.}
\label{fig10}
\end{figure}

In \cite{19} there is proposed a numerical algorithm for computing the homology groups of
$M_i $ by using a discrete version of Morse theory. Let us briefly expose the main constructions.
We consider the ``diagonal'' linear function
on $K$:
$$
f(x_1 ,x_2 ,x_3 ) = x_1 + x_2 + x_3 ,
$$
and the excursion sets
$$
M_i^a = \{\bar {x} \in M_i \vert f(\bar {x}) \le a\}.
$$
A critical point of $f$ is a vertex $v \in M_i $, i.e. an integer-valued point of the rectangular lattice in
$K$, such that when $a$ passes $af(v)$
the topology of $M_i^a$ changes. All combinatorial types of critical points
$v = (k_1 ,k_2 ,k_3 )$ are classified in terms of their elementary neighborhoods:
$$
N(v) = \{\bar {x} \in M\vert \vert x_i - k_i \vert \le 1,i = 1,2,3\}.
$$
A nondegenerate critical point has index $0$, $1$, or $2$ being the dimension of a cell that glued
to $M_i^a $ when $a$ passes the critical level. Moreover, there is a degenerate critical point, the ``monkey saddle,''
such that two $1$-dimensional cells are glued during passing the corresponding critical level.
In Fig. \ref{fig9} and Fig. \ref{fig10}
there are exposed the classical critical points:
the saddle defined by the equation $f(x,y) = x^2 - y^2)$ and the ``monkey saddle,'' defined by the equation
$f(x,y) = x^3 - xy^2$, and also their discrete analogs.

In \cite{19} it is constructed a chain complex
$$
C_2 (M_i ) \to C_1 (M_i ) \to C_0 (M_i ),
$$
consisting of vector spaces $C_q(M_i)$ over $\Z_2$.
The basic vectors in $C_q (M_i )$ correspond to the critical points of index $q$ and, moreover, the
monkey saddle correspond to a pair of basic vectors from $C_1 (M_i )$: to each monkey saddle
$v$ we add a fictive vertex $v^\prime$ which lies above $v$ with respect to the level of $f$ and a fictive edge
$vv^\prime$ that joins $v$ and $v'$).
The horizontal arrows denote the differentials, i.e. linear operators
$\partial _2^i :C_2 (M_i ) \to C_1 (M_i )$  and  $\partial _1^i
:C_1 (M_i ) \to C_0 (M_i )$ such that $\Ker \partial^i_1 = \Im \partial^i_2$.
We have
$$
B_q = \Im \{C_{q+1} \to C_q\}, \ \ Z_q = \Ker \{C_q \to C_{q-1}\}, \ \ H_q = Z_q/B_q
$$
where we assume that $C_3=C_{-1}=0$.

The differentials are constructed explicitly \cite{19} and for a demonstrative example
we need only the following property:

If $v$ is a critical point of index $1$, then there is a pair of sequences of vertices
$L(v) = [(v,v_1^ - ,\ldots ,v_k^ - )$, $(v,v_1^+,\ldots, v_m^ + )]$ such
that they contain exactly two critical points of index $0$ which are
$v_k^-$ and $v_m^ + $ and every two consecutive points $v_i^ - ,v_{i + 1}^ - $ or
$v_i^+, v_{i + 1}^+$ are connected by a negative edge, i.e.
such an edge that the value of $f$ at its end is less that at its
starting point. Then $\partial _1 (v) = v_k^ - + v_m^ + $.

Our task is to construct the homomorphisms $\varphi_q^i: C_q (M_i) \to C_q
(M_{i + 1}) $ compatible with differentials. After that, as explained
in the previous section, we can calculate the persistent homology and construct the
barcodes that reflect the dynamics of change of the topological structure of a $ M_i $
as $i$ increases. Immediately we understand the arising difficulty: the natural inclusion
$ M_i \subset M_{i + 1} $ induces no
the natural homomorphism $\varphi_q^i: C_q (M_i) \to C_q (M_{i + 1})$
at the critical points, and so there are no natural homomorphisms
of homology groups induced by the embedding $ M_i \subset M_{i + 1}$. This difficulty is
overcomed by the use of the discrete gradient flow
similar to that which was introduced in \cite{19}.

First we define the gradient descent of an arbitrary graph $\Gamma$ formed by edges.
Namely we assume that $\Gamma^\prime$ is obtained by an elementary descent of $\Gamma$, if 1)
$ (\Gamma \setminus \Gamma^\prime) \cup (\Gamma^\prime \setminus \Gamma)$
is the boundary (possibly without vertices) of the elementary face, and 2) all the vertices of
$\Gamma^\prime \setminus \Gamma$ lie on the lower levels of $f$ than all the vertices of $\Gamma
\setminus \Gamma^\prime$ and $\Gamma^\prime
\setminus \Gamma$ is not empty. If $\Gamma^\prime$ is obtained from $\Gamma$
by a finite sequence of elementary descents and there is no
elementary descent for $\Gamma^\prime$, then we say that $\Gamma^\prime$ is obtained from
$\Gamma$ by gradient descent.

Next we assume that we have a pair of three-dimensional bodies $M \subset N$,
consisting of elementary cubes. It suffices to construct homomorphisms of chain groups
for such pairs. Let $M_q$ and $N_q$ be the sets of critical points, of index $q$,
of $f$ in $M$ and in $N$. Let $v \in M_0 $. Given $v \in N_0$
put $\varphi_0 (v) = v$. Otherwise, there is a
negative edge $e_1$, in $ N $, starting at $v$, and let $v_1$ be its another end.
If $v_1 \in N_0$, then we put $\varphi_0
(v) = v_1$, and etc. We obtain an iterative process that results in
the chain $\Phi(v) = (v, v_1, \ldots, v_k)$, where $v \in M_0 $,
$v_k \in N_0 $, $ v_i \notin N_0 $ for $ 1 \le i  < k $, and all edges
$ [v_i, v_{i+ 1}] $ are negative. We put $\varphi _0(v) = v_k$.

Let $v \in M_1$. Let us construct $ L_M (v) = [(v, v_1^-, \ldots,
v_k^-), (v, v_1^+, \ldots, v_m^+)] $. To each of the sequences from $L_M (v)$,
we add the sequence $\Phi(v_k^-)$ or $\Phi
(v_m^+) $, respectively, and obtain a new pair of sequences
$[(v, \ldots, v_k^-, \ldots, v_p^-)$, $(v, \ldots,$ $v_m^+, \ldots, v_q^+)]$.
Let us construct the graph $\Gamma$ consisting of the edges
$[v_i ^ -, v_ {i + 1} ^ -] $,
$[v_i^+, v_ {i + 1} ^ +]$, $ 1 \leq i \leq k-1$, and
$[v, v_1^-]$, $ [v, v_1^+]$ and let $\Gamma^\prime $ be
obtained from $\Gamma$ by the gradient descent. Obviously,
the vertices $v_p^-$ and $v_q^+ \in N_0$ and their constituent edges cannot
down below. Therefore there exists a path $\gamma^\prime$ in $\Gamma^\prime$ which
connects $v_p^-$ and $v_q^+$. Let us consider all
critical points $w_1, \ldots, w_l$ of index $1$ in $\gamma^\prime$ and put
$\varphi_1(v) = \sum \limits_{i = 1}^l w_i$.

We have  $\varphi _0 (\partial _1 (v)) = \varphi _0
(v_k^ - + v_m^ + ) = v_p^ - + v_q^ + $ for $v \in M_1 $. But $\partial _1
(\varphi _1 (v)) = \partial _1 (\sum\limits_{i = 1}^l {w_i } )$.
Since the expansions for $\partial_1(w_i)$ and for $\partial_1(w_{i+1})$  have a common
component that is a critical point of index $0$ and the field of coefficients is of characteristic $2$,
$\partial_1(\varphi_1(v)) = v_p^- + v_q^+$.
Hence $\varphi _0 (\partial _1 (v)) = \partial _1 (\varphi _1 (v))$, i.e. the differentials commute
with the homomorphisms og homology groups induced by the embeddings.

The commutative diagram
$$
\begin{array}{ccccc}
 & C_1(M) & \stackrel{\varphi_1}{\longrightarrow} & C_1(N) & \\
 \partial_1 & \downarrow & & \downarrow & \partial_1 \\
 & C_0(M) & \stackrel{\varphi_0}{\longrightarrow} & C_0(N) &
 \end{array}
$$
enables us to compute the persistent $0$-homology.

To compute the persistent $2$-homology we have to use duality
\cite{19} and to compute the persistent $0$-homology for the dual space.
Namely, let us consider the three-dimensional body
$M^\prime = K\setminus M$, the complement to $M$, and the function $h = -f$.
Clearly the critical points of indices $0$, $1$, and $2$ of $h$ coincide with
the critical points of indices $2$, $1$, and $0$ of $f$ and so we may reduce
the computation of $2$-homology and $2$-barcodes of $M$ to the computation of
$0$-homology and $0$-barcodes of $M^\prime$.

\begin{figure}[htbp]
\centerline{\includegraphics[width=5.00in,height=3.00in]{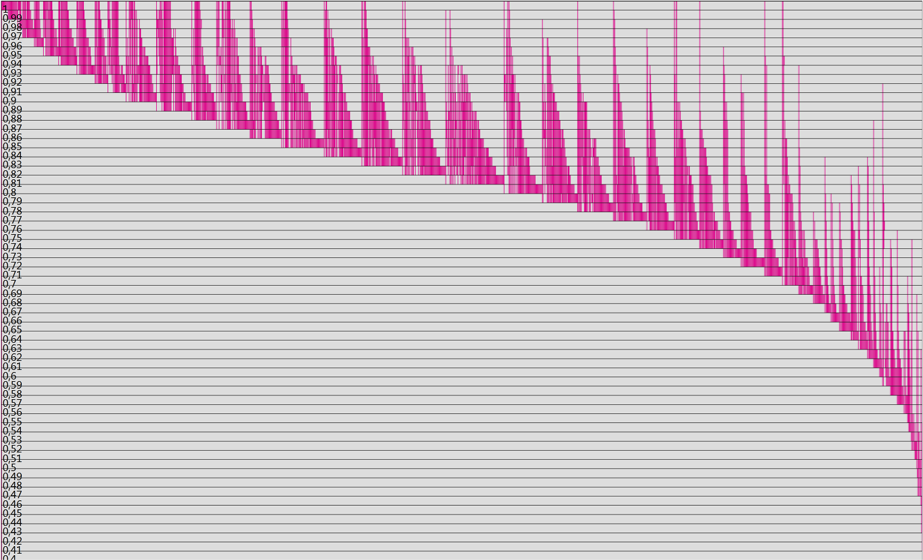}}
\vskip0.2cm
\centerline{\includegraphics[width=5.00in,height=3.00in]{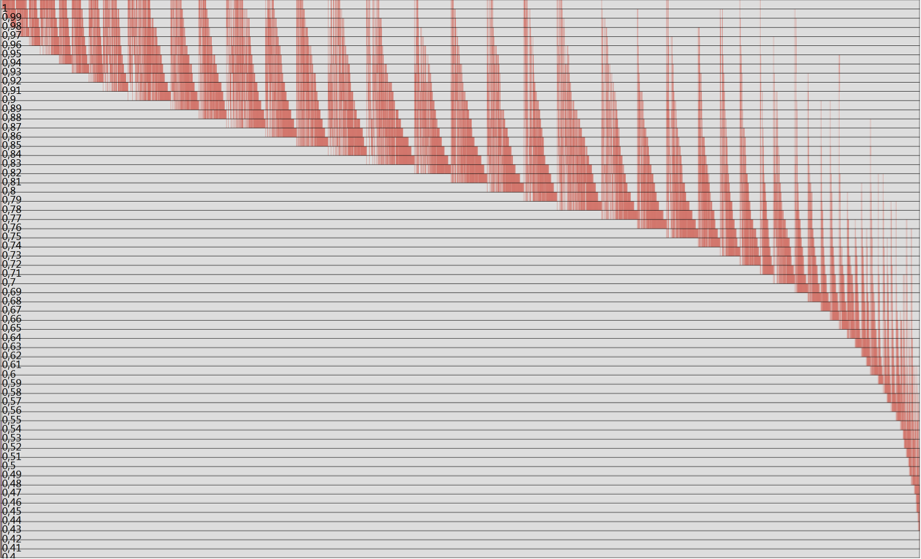}}
\caption{2-barcodes of the realizations obtained
by SPECTRAL (above)
and SGS (below). The excursion level values are plotted along the vertical.}
\label{fig11}
\end{figure}

In Fig. \ref{fig11} we present the barcodes of persistent $2$-homology of reservoirs
obtained by SPECTRAL (above) and SGS (below).

\newpage



\begin{thebibliography}{MMM}

\bibitem{1}
Baikov, V.A., Bochkov, A.S., and Yakovlev, A.A.:
The heterogeneuity of the Priobskoye field geological
modeling and simulation. Oil Industry (2011), N. 5, 50--54.
[Russian]

\bibitem{2}
Baikov, V.A., and Yakovlev, A.A.:
Reproduction of geological heterogeneity in geological and hydrodynamic models.
Rosneft Scientific and Technical Bulletin (2010), N. 2, 13--15.
[Russian]

\bibitem{3}
Baikov, V.A., Bezrukov, A.V., Bikbulatov, S.M., Emchenko, O.V., Mukharlyamov, A.R.,
Suleimanov, D.D., and Usmanov, T.S.:
A use of normal well development data for elimination of geostatistical modeling uncertainties.
Oil Industry (2009), N. 11, 16--19.
[Russian]

\bibitem{4}
Baikov, V.A., Emchenko, O.V., Roschektaev, A.P., and Yakovlev, A.A.:
Geological multifactor simulation exemplified by the Priobskoye field.
CKR Rosnedra Bulletin (2010), N. 1, 27--34.
[Russian]

\bibitem{5}
Edelsbrunner, H.:
Geometry and Topology for Mesh Generation. Cambridge
University Press, Cambridge, 2001.

\bibitem{6}
Kaczynski, T., Mischaikow, K., and Mrozek, M.:
Computational Homology.
Appl. Math. Sci. Series {\bf 157}, Springer-Verlag, New York, 2004.

\bibitem{7}
Zomorodian, A.J.:
Topology for Computing. Cambridge University Press, Cambridge, 2005.

\bibitem{8}
Edelsbrunner, H., Letscher, D., and Zomorodian, A.:
Topological persistence and simplification.
Discrete Comput. Geom. {\bf 28} (2002), 511--533.

\bibitem{9}
Zomorodian, A., and Carlsson, G.:
Computing persistent homology.
Discrete Comput. Geom. {\bf 33} (2005), 249--274.

\bibitem{10}
Cohen-Steiner, D., Edelsbrunner, H., and Harer, J.:
Stability of persistence diagrams, in: Proc. 21st Sympos. Comput. Geom. (2005), 263--271.

\bibitem{12}
Bubenik P., and Kim, P.T.:
A statistical approach to persistent homology.
Homology, Homotopy and Applications {\bf 9} (2007), 337--362.

\bibitem{16}
Edelsbrunner, H., and Harer, J.:
Persistent homology --- a survey. In:
Surveys on discrete and computational geometry. Contemp. Math. {\bf 453},
Amer. Math. Soc., Providence, RI, 2008, pp. 257--282.

\bibitem{11}
Ghrist R.. Barcodes: The persistent topology of data. Bull. Amer. Math.
Soc. {\bf 45} (2008), 61--75.

\bibitem{14}
Carlsson G.:
Topology and data.
Bull. Amer. Math. Soc. {\bf 46} (2009), 255--308.

\bibitem{15}
Carlsson, G., and Zomorodian, A.:
The theory of multidimensional persistence.
Discrete Comput. Geom. {\bf 42} (2009), 71--93.

\bibitem{17}
Adler, R.J., Bobrowski, O., Borman, M.S., Subag, E., and Weinberger, S.:
Persistent Homology for Random Fields and Complexes.
In: Borrowing strength: theory powering applications -- a Festschrift for Lawrence D. Brown, 124--143,
Inst. Math. Stat. Collect., {\bf 6}, Inst. Math. Statist., Beachwood, OH, 2010.

\bibitem{18}
Adler, R.J., and Taylor, J.E.:
Random Fields and Geometry. Springer Monographs
in Mathematics, Springer, New York, 2007.

\bibitem{19}
Bazaikin, Ya.V., and Taimanov, I.A.:
On a numerical algorithm for computing topological characteristics of three-dimensional bodies.
J. of Comp. Math. and Math. Phys. 2013 (to appear) [Russian];
arXiv:1302.3669.

\bibitem{20}
Deutsch, C.V., and Journel, A.G.:
GSLIB, Geostatistical Software Library and User's Guide. Oxford University Press, New York, 1992.



\bibitem{21}
Baikov, V.A., Bakirov, N.K., and Yakovlev, A.A.:
New approaches in geostatistical modeling theory.
Vestnik UGATU {\bf 37}:2 (2010), 209--215.
[Russian]

\bibitem{22}
Baikov, V.A., Bakirov, N.K., and Yakovlev, A.A.:
New approaches to the geological and hydrodynamic modeling.
Oil Industry (2010), N. 9, 56--59.
[Russian]

\bibitem{23}
Seifert, H., and Threlfall, W.:
Variationsrechnung im Grossen, AMS Chelsea Publishing, Providence, R.I., 1971.

\bibitem{24}
Dubrovin, B.A., Fomenko, A.T., and Novikov, S.P.:
Modern Geometry - Methods and Applications: Part III: Introduction to Homology Theory.
Graduate Texts in Mathematics, {\bf 124}. Springer, New York, 1990.

\end{thebibliography}
\end{document}